%% file: main.tex
\documentclass[preprint,journal]{vgtc}                

\ifpdf
  \pdfoutput=1\relax                   
  \usepackage{graphicx}                
  \DeclareGraphicsExtensions{.pdf,.png,.jpg,.jpeg} 
\else
  \usepackage{graphicx}                
  \DeclareGraphicsExtensions{.eps}     
\fi%

\graphicspath{{figures/}{pictures/}{images/}{./}} 

\usepackage{microtype}                 
\PassOptionsToPackage{warn}{textcomp}  
\usepackage{textcomp}                  
\usepackage{mathptmx}                  
\usepackage{times}                     
\usepackage{cite}                      
\usepackage{tabu}                      
\usepackage{booktabs}                  
\usepackage{comment}                   
\usepackage{amsmath}
\usepackage{diagbox}
\usepackage{listings}
\usepackage{soul}
\usepackage{color}
\usepackage[dvipsnames]{xcolor}
\usepackage{amssymb}

\onlineid{1122}

\vgtccategory{Research}
\vgtcpapertype{algorithm/technique}

\title{Selection Bias Tracking and Detailed Subset Comparison for High-Dimensional Data}

\author{David Borland, Wenyuan Wang, Jonathan Zhang, Joshua Shrestha, and David Gotz}
\authorfooter{
\item
 David Borland is with RENCI at the University of North Carolina at Chapel Hill. E-mail: borland@renci.org.
\item
 Wenyuan Wang is with the School of Information and Library Science at the University of North Carolina at Chapel Hill. E-mail: vaapad@live.unc.edu.
\item
 Jonathan Zhang is with the Dept. of Biostatistics at the University of North Carolina at Chapel Hill. E-mail: jzhang42@live.unc.edu.
\item
 Joshua Shrestha is with the Dept. of Computer Science at the University of North Carolina at Chapel Hill. E-mail: prayash@live.unc.edu.
\item
 David Gotz is with the School of Information and Library Science at the University of North Carolina at Chapel Hill. E-mail: gotz@unc.edu.
}

\shortauthortitle{Biv \MakeLowercase{\textit{et al.}}: Global Illumination for Fun and Profit}

\abstract{
The collection of large, complex datasets has become common across a wide variety of domains. Visual analytics tools increasingly play a key role in exploring and answering complex questions about these large datasets. However, many visualizations are not designed to concurrently visualize the large number of dimensions present in complex datasets (e.g. tens of thousands of distinct codes in an electronic health record system). This fact, combined with the ability of many visual analytics systems to enable rapid, ad-hoc specification of groups, or cohorts, of individuals based on a small subset of visualized dimensions, leads to the possibility of introducing selection bias--when the user creates a cohort based on a specified set of dimensions, differences across many other unseen dimensions may also be introduced. These unintended side effects may result in the cohort no longer being representative of the larger population intended to be studied, which can negatively affect the validity of subsequent analyses. We present techniques for selection bias tracking and visualization that can be incorporated into high-dimensional exploratory visual analytics systems, with a focus on medical data with existing data hierarchies. These techniques include: (1) tree-based cohort provenance and visualization, including a user-specified baseline cohort that all other cohorts are compared against, and visual encoding of cohort ``drift'', which indicates where selection bias may have occurred, and (2) a set of visualizations, including a novel icicle-plot based visualization, to compare in detail the per-dimension differences between the baseline and a user-specified focus cohort. These techniques are integrated into a medical temporal event sequence visual analytics tool. We present example use cases and report findings from domain expert user interviews.  
} 


\CCScatlist{ 
 \CCScat{K.6.1}{Management of Computing and Information Systems}%
{Project and People Management}{Life Cycle};
 \CCScat{K.7.m}{The Computing Profession}{Miscellaneous}{Ethics}
}

\teaser{
  \centering
  \includegraphics[width=0.92\linewidth]{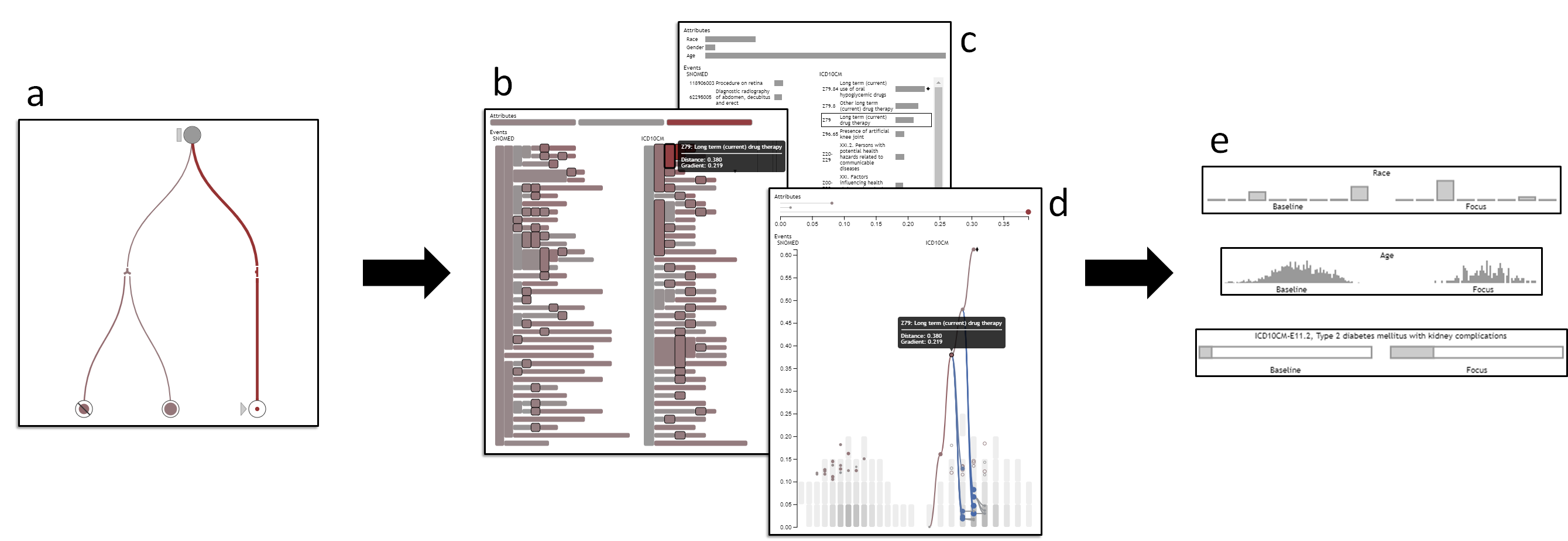}
  \vspace{-0.35cm}
  \caption{Exploratory cohort selection in high-dimensional datasets can lead to selection bias--unintended side-effects in variable distributions--that may go unnoticed by the user. Our selection bias tracking system and detailed cohort comparison visualizations, deployed in a medical temporal event sequence visual analytics tool, include (a) a cohort provenance tree to keep track of created cohorts and indicate when selection bias may have occurred, (b-d) a suite of high-dimensional cohort comparison visualizations that employ hierarchical aggregation to display the differences between two cohorts in detail, and (e) data-type dependent comparisons of individual variable distributions.}
	\label{fig:teaser}
}

\vgtcinsertpkg

\begin{document}

\input{sections/intro.tex}
\input{sections/related.tex}

\input{sections/system.tex}
\input{sections/requirements.tex}
\input{sections/bias.tex}

\input{sections/design.tex}
\input{sections/evaluation.tex}
\input{sections/conclusion.tex}

\acknowledgments{
The research reported in this article was supported in part by a grant from the National Science Foundation (\#1704018).}

\bibliographystyle{abbrv-doi}

\bibliography{VAST_2019_Cohort_Comparison.bib}
\end{document}

%% file: sections/intro.tex
\firstsection{Introduction}
\label{sec:Introduction}

\maketitle

%
%
%
%

The collection of large, complex datasets has become common across a wide variety of domains, such as advertising, security, and healthcare. In addition to analytical approaches such as statistics, data mining, and machine learning, visual analytics increasingly plays a key role in exploring and answering complex questions about such large datasets to support precision, evidence-based decision making \cite{thomas_illuminating_2005,keim_big-data_2013}.

Two distinct challenges encountered with large datasets are data \textit{volume} and data \textit{complexity}. Volume typically refers to the number of records in a given dataset, e.g. the number of patients in an electronic health record (EHR) system. Complexity, meanwhile, reflects the number of dimensions, e.g. the various demographic, diagnosis, procedure, lab, and medication data stored for each patient in an EHR. For many analytical tasks increased volume can be managed with improved processing power. Moreover, many visualizations scale well with increasing volume because they depict aggregate values---a bar chart of a binary variable works equally well with ten records as with 1 billion records. Data complexity, however, presents a more fundamental challenge. For example, many EHRs use coding systems that can contain hundreds of thousands of unique codes to represent different diagnoses, procedures, medications, etc. (e.g. \cite{noauthor_icd_2018,spackman_snomed_1997}). Although many high-dimensional visualization techniques exist (e.g. \cite{liu_visualizing_2017,fayyad_information_2001} provide reviews), they can typically only show a relatively small subset of the dimensions represented in such complex datasets at any given time, via techniques such as dimension selection or projection. 

At the same time, a wide variety of visual analytics applications have successfully employed Shneiderman's mantra of ``overview first, zoom and filter, then details-on-demand'' \cite{shneiderman_eyes_1996} to help manage the complexity of high-dimensional datasets. Indeed, filtering is a common step in many analytic workflows.

However, the ability of sophisticated visual analytics systems to enable rapid, ad-hoc filtering of complex datasets while simultaneously focusing the user's attention on a narrow subspace of the overall dataset at any given point in time can create a situation ripe for issues such as \textit{selection bias}. Selection bias occurs when the users selects a sample for analysis in such a way that the sample is not representative of the larger group that is intended to be analyzed \cite{hernan_structural_2004}. For example, in the medical domain, certain medications cannot be used together due to drug-drug interactions. As a result, filtering patient data on one drug during a medical analysis will result in a patient population with a much lower-than-typical rate of the second drug. This shift in the distribution of the second drug
may be an \textit{unintended side-effect} of the filtering operation, and may go unnoticed by the user, especially if data for the second drug are not currently shown in the visualization. 

More generally, when a user applies filters to create data subsets based on a small set of dimensions,
there may be large shifts in the distributions of other ``unseen'' correlated dimensions. 
If such side-effects go unnoticed, they could threaten the validity of subsequent analyses. 
In the medical domain, such side-effects can be especially problematic because there are many inter-related dimensions, and selection bias could lead to incorrect evidence for medical decision making.


Contextual visualization methods \cite{borland_contextual_2018} are one proposed approach for addressing selection bias and related challenges. One such method, deployed in a medical application to address selection bias, is adaptive contextualization \cite{gotz_adaptive_2016,gotz_adaptive_2017}. In this approach, subsets of patients, or cohorts, are interactively generated by the user of a visual analytics interface depicting medical event sequences. The system keeps track of each generated cohort, and computes a distance measure indicating to what degree the variable distribution of each cohort has ``drifted'' from a baseline cohort, which may indicate unintended selection bias. The user can inspect a list of event types to see the degree of drift for each individual variable. Although shown to be effective, the system exhibits two critical limitations: (1) it does not enable key types of bias comparisons required by various fields, including medical cohort analysis \cite{moher_consort_2010}, because it only considers a linear sequence of selection steps, and (2) it does not account for important variable interactions, including hierarchical relationships, focusing instead on independent univariate measures.

This paper builds on this earlier work, introducing a new approach that addresses these limitations. 
Selection bias is a known issue in various medical domains (e.g. \cite{hernan_structural_2004,torner_proposed_2010,torner_method_2011,nohr_how_2018}), and we focus our work on cohort selection from medical event sequence data, taking advantage of existing medical data hierarchies. Given a visual analytics interface that enables the user to create multiple cohorts of patients via various filter paths from a root dataset, the primary goals of the work in this paper are as follows: (1) Enable visualization of the \textit{created cohorts and the steps taken to create them}. (2) Enable the user to identify the \textit{degree to which the variable distribution for each cohort has drifted} from a user-specified baseline--an indicator of potential selection bias. (3) Enable the user to identify \textit{which filter operations contribute the most} to the potential selection bias for any cohort. (4) For a user-selected focus data cohort, enable the user to investigate in-depth the \textit{drift for each dimension} compared to the current baseline, prioritizing dimensions which have drifted most drastically. (5) Leverage \textit{hierarchical relationships} between dimensions to better communicate areas of drift within high-dimensional data. Accomplishing these goals will alert the user to any selection bias introduced in their current analysis, enabling them to correct for such bias in subsequent analyses.

The key research contributions presented in furtherance of these goals include:

\vspace{-0.1cm}   
\begin{itemize}
\vspace{-0.1cm}   
    \item A tree-based cohort provenance visualization, showing the full non-linear selection process from an initial query result to one or more final cohorts, along with the amount of selection bias introduced at each step. The user can interactively select a baseline cohort against which all cohorts are compared, and a focus cohort for detailed comparison with the baseline.
\vspace{-0.15cm}   
    \item A set of visualizations for communicating selection bias between pairs of cohorts from a high-dimensional dataset, including a novel visualization technique adapted from icicle plots. A user-specified aggregation level is incorporated into each visualization to scale effectively for high-dimensional datasets, while prioritizing salient dimensions with respect to selection bias.
\vspace{-0.15cm}   
    
    
    
    
    \item Integration of these, and other relevant cohort comparison visualizations, within \textit{Cadence}, a temporal event sequence visual analytics and cohort selection tool.
\vspace{-0.1cm}   
\end{itemize}

This paper describes in detail the contributions listed above, presents example use cases, and reports findings from medical domain expert interviews. 

%% file: sections/related.tex
\section{Related Work}

The section provides a brief overview of related work that is most relevant to the contributions of this paper: hierarchical visualization, visual comparison, bias, and medical cohort management.

\subsection{Hierarchical Visualization}

Both the cohort provenance (Section \ref{sec:CohortTree}) and two of the detailed cohort comparison visualizations (Sections \ref{sec:IciclePlot} and \ref{sec:DotPlot}) presented in this paper involve the visualization of hierarchies. A hierarchy consists of a set of nodes $N$ and directed edges $E$. A single \textit{root} node has only outgoing edges; all other nodes must have one incoming edge, from its \textit{parent}, and zero or more outgoing edges, to its \textit{children}.
Hierarchies are used across a wide range of domains to organize data, and hierarchical aggregation is often used to manage visual complexity in data visualization \cite{elmqvist_hierarchical_2010,zinsmaier_interactive_2012}. 
The various visualization approaches developed to communicate hierarchical structures are often classified as node-link representations or implicit/space-filling techniques \cite{schulz_design_2011}.

Node-link diagrams represent each node in the hierarchy as a glyph, and each hierarchical relationship as a mark (e.g. line, curve, etc.) connecting each child node glyph to its parent. Layout strategies for node-link diagrams include phylogenetic trees (e.g., \cite{swofford_phylogenetic_1996}) and dendrograms, which draw all leaf nodes at the same depth and are often used for displaying hierarchical clustering output (e.g., \cite{rokach_clustering_2005}). In contrast, layouts such as tidy trees \cite{reingold_tidier_1981} draw each level in the tree at the same depth, potentially resulting in a ``ragged'' appearance for leaf nodes. Our cohort provenance tree uses a node-link diagram to enable the encoding of information along each link in the tree, with a tidy-tree layout to emphasize the sequences of steps taken to form each cohort.

Implicit techniques include treemaps \cite{johnson_tree-maps:_1991,shneiderman_tree_1992} and icicle plots \cite{kruskal_icicle_1983}. These techniques do not draw parent-child relationships directly, instead encoding them using enclosure or adjacency. As a result, they offer compact representations and are often suitable for the visualization of large hierarchies.  However, the ability to encode information for each link is hindered. For our detailed cohort comparison visualizations we explore the use of implicit (Section \ref{sec:IciclePlot}) and node-link (Section \ref{sec:DotPlot}) representations, incorporating a user-specified aggregation level while prioritizing salient features in the hierarchy.


\subsection{Visual Comparison}
Comparison is a fundamental process for many visualization tasks, and a variety of visual comparison systems and approaches have been developed to meet specific needs. In light of such diversity, Gleicher et al. developed a taxonomy of comparative visualization techniques: (1) juxtaposition, by showing objects separately, (2) superposition, by overlaying objects in the same space, and (3) explicit encoding, by directly representing the relationships between objects \cite{gleicher_visual_2011}. 

The cohort provenance visualization described in this paper adopts a tree structure that is inspired in part by CONSORT diagrams (Section~\ref{sec:CohortManagement}). We therefore employ an explicit encoding of comparison between nodes in the tree (representing cohort drift) that is overlaid onto the tree structure.

Various approaches have been developed for the visual comparison of hierarchical data, including static topology over time (e.g.  \cite{wattenberg_designing_2006,baur_touchwave:_2012,cuenca_multistream:_2018}), interactive tree comparison (e.g. \cite{isenberg_interactive_2007,bremm_interactive_2011}), stable treemap layouts for improved juxtaposition comparisons (e.g. \cite{tu_visualizing_2007,sud_fast_2010,tak_enhanced_2013,hahn_visualization_2014,hees_stable_2015,sondag_stable_2018}), and dynamic topology (e.g. \cite{lukasczyk_nested_2017,kopp_temporal_2019}). Cui et al. employ dynamic simplification of hierarchies evolving over time, with the ability to interactively zoom to lower levels of detail \cite{cui_how_2014}. Our detailed cohort comparison visualizations adopt a similar strategy, employing user-controlled hierarchical aggregation and interactive level-of-detail exploration. However, our focus is on conveying where in the hierarchical data structure two cohorts are most different. Both the split icicle plot (Section~\ref{sec:IciclePlot}) and dot plot (Section~\ref{sec:DotPlot}) visualization designs explicitly encode the differences between each cohort, conveyed via a shared hierarchical structure.

\begin{figure}
    \centering
    \includegraphics[width=0.45\textwidth]{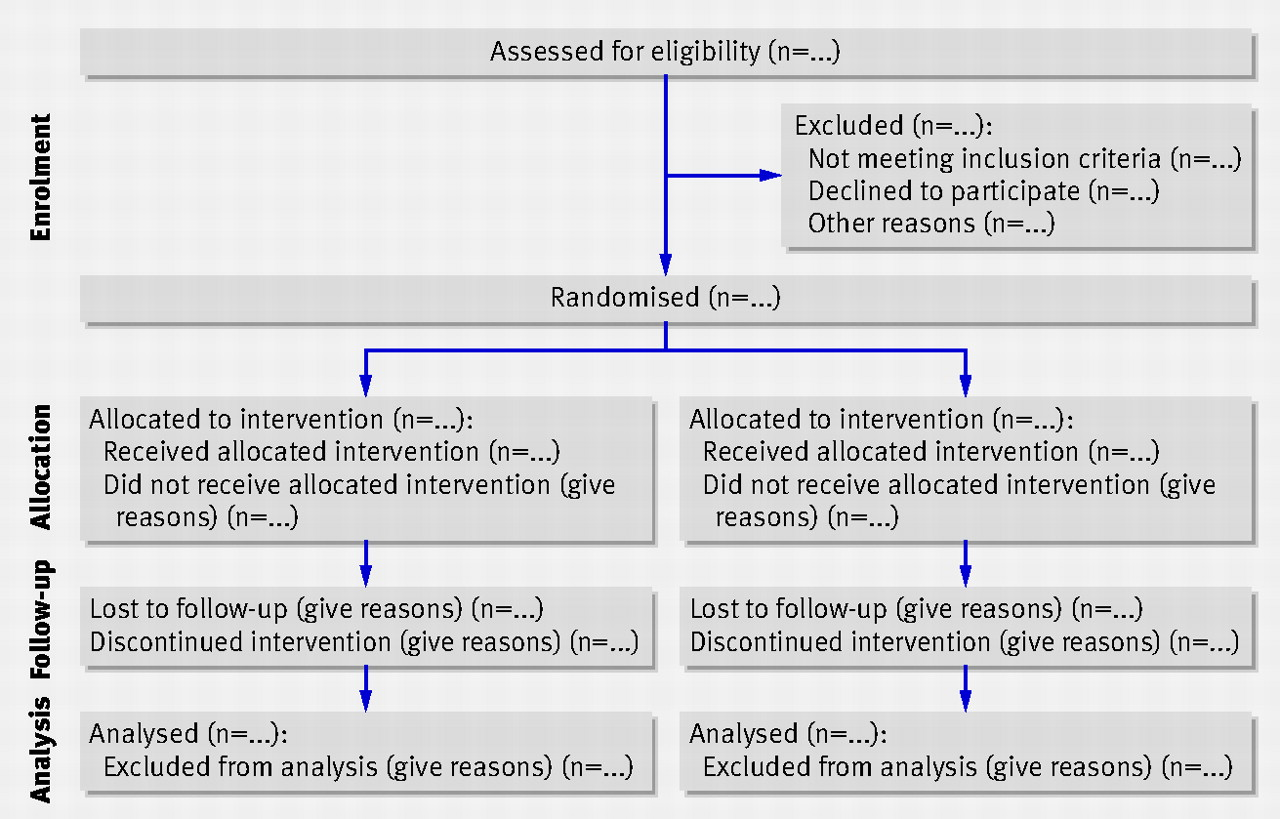}
    \caption{CONSORT flow diagram for a two-group randomized trial \cite{moher_consort_2010}.}
    \label{fig:flow_diagram}
\end{figure}

\subsection{Bias}

Bias is the (unfair) disproportionate weighting in favor of or against one thing compared to another. Various forms of bias have been recognized, including cognitive and statistical biases. Although domain expertise represents a form of bias that can help an analyst complete a task, e.g. via heuristic short cuts for filtering out unnecessary information \cite{gigerenzer_why_2008}, bias can also negatively affect the integrity of an analytical process and the validity of analytical results. For this reason, bias has increasingly become a topic of study across a variety of disciplines, including machine learning 
and visual analytics (e.g. \cite{dimara_task-based_2018,ellis_four_2018,ellis_cognitive_2018}). Much of the focus in the visualization community has focused on characterizing and addressing cognitive biases \cite{wall_warning_2017,cho_anchoring_2017,valdez_priming_2018}. Expanding on prior work in adaptive contextualization \cite{gotz_adaptive_2016,gotz_adaptive_2017}, the work presented in the paper focuses on the issue of selection bias--when individuals are selected for analysis in such a way that the sample is not representative of the population intended to be analyzed--an issue exacerbated by the interactive analysis of large, complex datasets, and of particular concern within the medical domain. 
 
\subsection{Medical Patient Cohort Management}
\label{sec:CohortManagement}

The randomized control trial (RCT) is a widely used experimental design in fields such as epidemiology.  The Consolidated Standards of Reporting Trials (CONSORT) statement \cite{schulz_consort_2010, moher_consort_2010} has been introduced for rigorous reporting of RCTs to avoid biased results (e.g. \cite{schulz_empirical_1995}). The CONSORT flow diagram presents how RCTs progress through various phases (Figure \ref{fig:flow_diagram}). The tree-structure of the diagram shows the construction of cohorts with different inclusion and exclusion criteria. Interactive cohort selection tools for medical research, such as i2b2 \cite{murphy_serving_2010,harris_i2b2t2:_2016}, enable researchers to query EHR databases and define cohorts based on various criteria. However, they do not typically include visual comparisons of constructed cohorts. Given the familiarity of CONSORT diagrams to the medical target audience for the techniques presented in this paper, they motivate in part the visual design adopted in the cohort provenance visualization (Section \ref{sec:CohortTree}). This influence is reflected in the tree-based structure, the notions of inclusion and exclusion criteria, and the concepts of included and excluded cohorts.

Within the visualization community, a number of approaches have been developed to support patient cohort analysis (Rind et al. provide a review \cite{rind_interactive_2013}), including various approaches for temporal event sequences (e.g. \cite{gotz_decisionflow:_2014,malik_cohort_2015}), and cross-sectional phenotype studies \cite{glueck_phenostacks:_2017}. However, these systems are not designed to explicitly track and visualize selection bias.


%% file: sections/system.tex
\section{System Overview}

\begin{figure}
    \centering
    \includegraphics[width=0.5\textwidth]{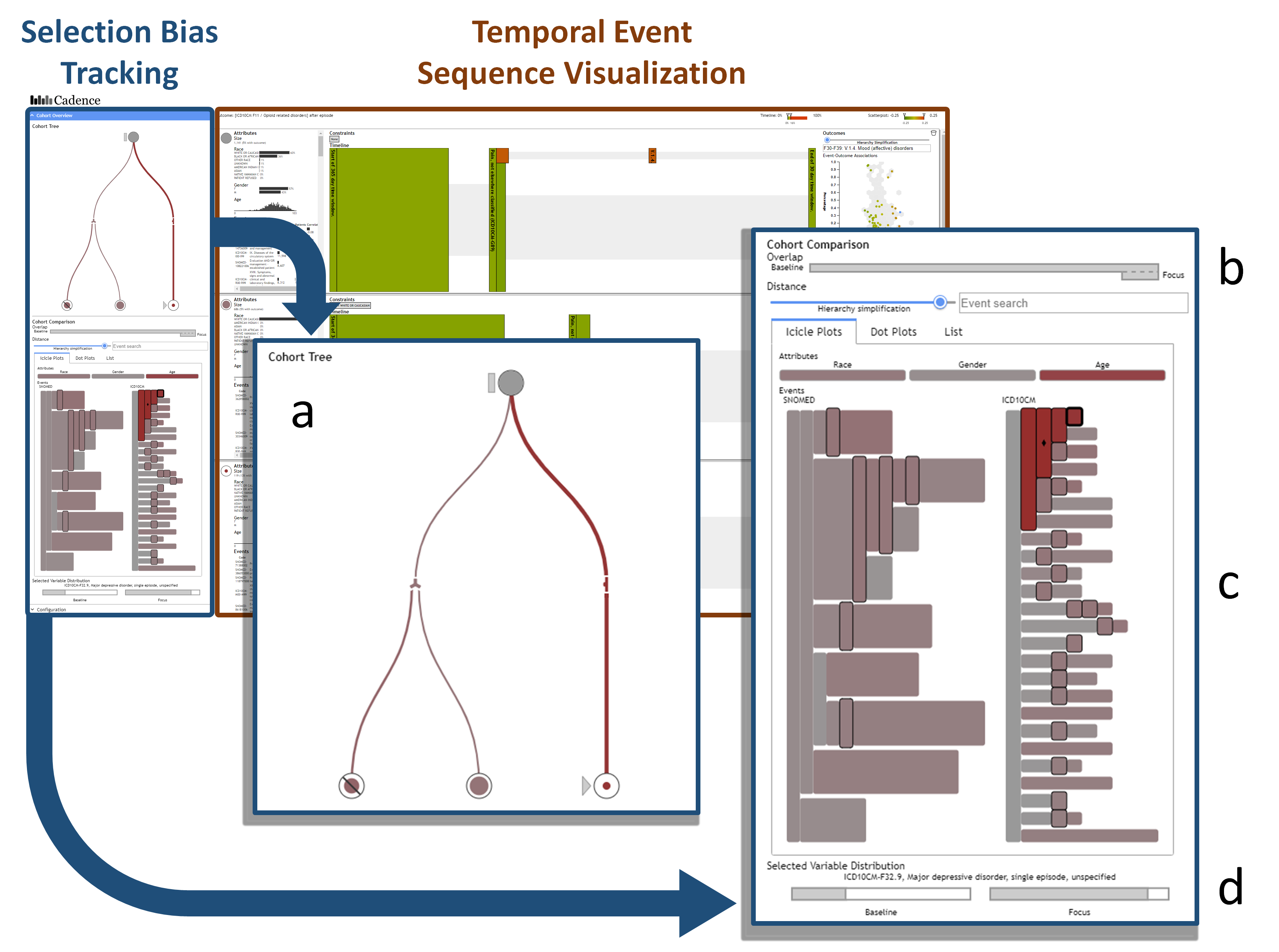}
    \caption{Overview of the \textit{Cadence} visual analytics tool, with \textcolor{NavyBlue}{selection bias tracking} (the focus of this paper) on the left, and \textcolor{Bittersweet}{temporal event sequence visualization} on the right.
    Closeup images of the selection bias tracking components show the (a) cohort provenance tree, (b) cohort overlap, (c) detailed cohort distance, and (d) selected variable distribution visualizations.}
    \label{fig:cadence}
\end{figure}

The selection bias tracking and detailed cohort comparison methods described in this paper are integrated into \textit{Cadence}, a medical temporal event sequence visual analytics tool (Figure \ref{fig:cadence}). \textit{Cadence} enables the selection of multiple cohorts of patients from an initial query result by filtering based on demographic attributes (\textit{Race}, \textit{Gender}, and \textit{Age}) and sequences of temporal events (\textit{Diagnoses} and \textit{Procedures}). The precise nature of the methods implemented for visualizing temporal events and specifying cohorts is beyond the scope of this paper, and described elsewhere \cite{gotz_visual_2020}. 
Instead, treating the cohort selection mechanisms as a black box, the bias tracking system that is the focus of this paper takes as its input (1) a set of cohorts $C=\{c_i\}$, where each cohort $c_i$ is defined by a set of patients (each with their own respective attributes and events), and (2) a set of operators $O=\{o_i\}$, where each operator $o_i$ corresponds to a filter (based on an attribute value or the presence or absence of an event) that defines the transformation of a parent cohort $c_j$ to a child cohort $c_k$. 

\subsection{Data Description}
\label{sec:DataDescription}

In \textit{Cadence}, medical data is represented using standardized coding systems including ICD-10-CM for diagnoses \cite{noauthor_icd_2018} and SNOMED-CT for procedures \cite{spackman_snomed_1997}.  Both coding systems are hierarchical in structure, with ICD-10-CM containing over 70,000 distinct codes and SNOMED-CT over 300,000 distinct codes (of which a subset corresponds to medical procedures).  Typically, data from EHR systems can contain codes from various levels of the coding hierarchy. For example, one patient may be diagnosed with the ICD-10-CM code \textit{150: Heart Failure}, whereas another may be diagnosed with a more specific code, such as \textit{150.32: Chronic diastolic (congestive) heart failure}. For each patient, medical event types can be viewed as binary variables: \textit{present} (recorded in the patient's data) or \textit{absent} (not recorded in the patient's data). Moreover, if a given code is present for a patient, all ancestors of that code in the code hierarchy are also considered present.  For example, a patient with \textit{I50.32} would also be considered to have the more generic \textit{I50} diagnosis code.  Finally, the data are relatively sparse, with some codes used frequently, others rarely, and many not at all within a given cohort.

To provide a sense for the complexity of the data in this domain, Table \ref{tab:querystats} summarizes statistics for a dozen queries run against real-world medical data in \textit{Cadence}. The smallest query result returned 3,594 distinct codes from the SNOMED-CT hierarchy and 8,159 from the ICD-10-CM hierarchy, whereas the largest contained 4,750 unique SNOMED-CT codes and over 10,000 unique ICD-10-CM codes. 
The large numbers of unique codes in a typical query result motivated the development of the hierarchical aggregation methods used for the split icicle plot (Section \ref{sec:IciclePlot}) and hierarchical dot plot visualizations (Section \ref{sec:DotPlot}) described later in this paper.

\begin{table}[t]
  \centering
        \small
      \begin{tabular}{|l|r|r|r|r|}
      \cline{3-5}
        \multicolumn{2}{l}{}
        & \multicolumn{3}{|c|}{\bf Distinct Event Types} \\
      \cline{2-5}
        \multicolumn{1}{c|}{}
        & \multicolumn{1}{c|}{\bf Patients}
        & \multicolumn{1}{c|}{\bf SNOMED-CT}
        & \multicolumn{1}{c|}{\bf ICD-10-CM}
        & \multicolumn{1}{c|}{\bf Total} \\
      \hline
    \bf Minimum & 1,732 & 3,594 & 8,159 & 11,753 \\
    \bf Average & 4,936 & 4,305 & 9,692 & 13,997 \\
    \bf Maximum & 8,360 & 4,750 & 10,626 & 15,376 \\
      \hline
      \end{tabular}
      \vspace{0.1cm}
    \caption{Statistics summarizing event data returned by 12 queries using \emph{Cadence}, each of which would form the initial cohort for an analysis.}
    \label{tab:querystats}
\end{table}

%% file: sections/requirements.tex
\section{Requirements for Selection Bias Tracking and Detailed Cohort Comparison}
\label{sec:Requirements}

Our system aims to alert the user to potential selection bias that can occur during rapid, ad-hoc cohort selection in high-dimensional datasets, leading to the following design requirements:

\begin{itemize}
    \item [\textbf{R1}] Provide an intuitive interface for keeping track of created cohorts and the steps taken to create them.
    \vspace{-0.20cm}
    \item [\textbf{R2}] Enable identification of which cohorts may be subject to selection bias.
    \vspace{-0.20cm} 
    \item [\textbf{R3}] Enable identification of which filter operations contribute the most to potential selection bias for a given cohort.
    \vspace{-0.20cm}
    \item [\textbf{R4}] Enable in-depth investigation of the distribution drift for each dimension between any two cohorts.
    \vspace{-0.20cm}
    \item [\textbf{R5}] Leverage hierarchical relationships between dimensions to better communicate areas of drift in high-dimensional data.
\end{itemize}

%% file: sections/bias.tex
\section{Measuring Selection Bias}
\label{sec:measuringbias}

A fundamental need for addressing the proposed requirements is the ability to measure potential selection bias. In the following sections we describe such a metric, and discuss specific considerations for analyzing and visualizing selection bias in the context of hierarchical data.

\subsection{Selection Bias Metric}
\label{sec:BiasMetric}

In order to detect selection bias, we must be able to quantify shifts in variable (attribute and event) distributions between cohorts. In a manner similar to adaptive contextualization \cite{gotz_adaptive_2016,gotz_adaptive_2017}, we use the Hellinger distance \cite{simpson_minimum_1987,pollard_users_2002}. However, we compute the distance at each level in the hierarchy for aggregate values, as discussed in Section \ref{sec:DataDescription}, rather than only considering each variable independently. The Hellinger distance is an established statistical metric that quantifies the similarity between two probability distributions. Its discrete form computes a distance between two discrete probability distributions $P = (p_1, ..., p_n)$ and $Q = (q_1, ..., q_n)$, where $n$ is the number of possible values for the variable (e.g. $n=2$ for a binary variable), as follows:
\begin{equation}
H(P,Q) = \sqrt{\frac{1}{2}\sum_{i=1}^{n}(\sqrt{p_i}-\sqrt{q_i})^2}
\label{eq:hellinger}
\end{equation}

Regardless of the value of {$n$}, $H=0$ when $P$ and $Q$ are identical, and $H=1$ when $P$ and $Q$ are maximally different. This characteristic makes the distance comparable when applied to heterogeneous variable types, and the discrete form of $H$ can be applied to categorical, ordinal, and ratio (with binning to discretize the distribution) variables.

When comparing two cohorts $c_{i}$ and $c_{j}$, $H$ is computed for all $m$ dimensions $d$ in the dataset. These individual distances are used in the detailed cohort comparison visualizations (Sections \ref{sec:IciclePlot}-\ref{sec:ListView}). However, since there can be thousands of individual distances, it is difficult to map data at this granularity onto the cohort provenance tree (Section \ref{sec:CohortTree}) for detecting potential selection bias. We therefore compute the average Hellinger distance between two cohorts $c_i$ and $c_j$ as:
\begin{equation}
H_{avg}(c_{i},c_{j}) = \frac{1}{m} \sum_{k=1}^{m}H(d_{ik},d_{jk})
\label{eq:hellinger_avg}
\end{equation}

This univariate metric is used in the cohort provenance tree as an indication of potential selection bias. The distances for each individual variable can then be investigated in more depth via the detailed cohort comparison techniques.

\subsection{Selection Bias in Hierarchical Data}
\label{sec:HierarchicalBias} 

\begin{figure}
    \centering
    \includegraphics[width=0.5\textwidth]{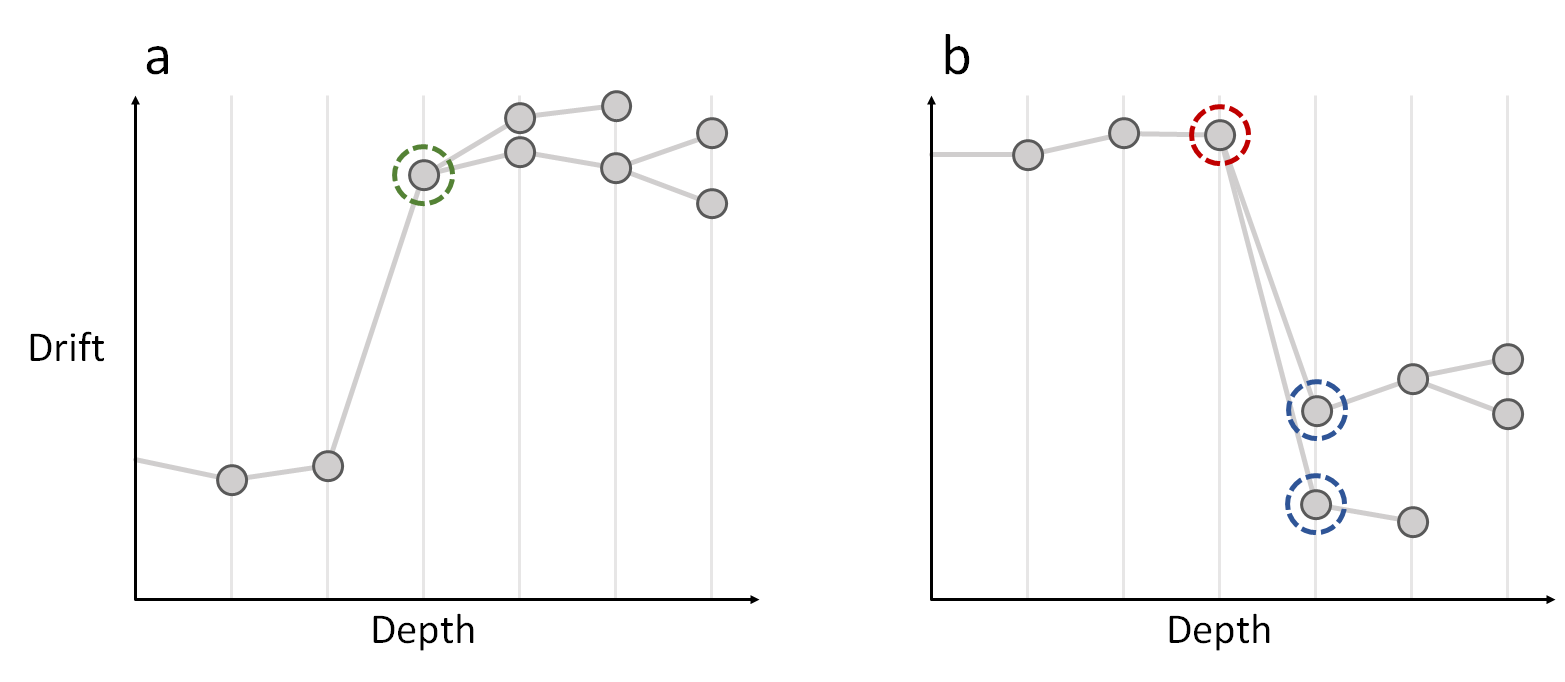}
    \caption{Examples of changing drift in a section of a hierarchy. Each dimension is a circle, with x-position the depth in the hierarchy and y-position the amount of drift. Links indicate parent-child relationships. (a) The \textcolor{OliveGreen}{green} node does not have the highest drift, but does \textit{differ} the most from its parent, indicating an area of the hierarchy where drift has increased. (b) Similarly, the circled \textcolor{Red}{red} and \textcolor{NavyBlue}{blue} nodes indicate an area where drift has decreased.} 
    \label{fig:hierarchy_difference}
\end{figure}

When comparing the distributions drifts between two cohorts for every dimension, it is important to convey which dimensions have drifted the most (measured by Equation \ref{eq:hellinger}), as has been considered in previous work \cite{gotz_adaptive_2016,gotz_adaptive_2017}. However, looking at each of these individual univariate distances in isolation does not provide information on \textit{where in the hierarchy areas of drift have been introduced}, nor on \textit{drift that emerges at higher levels of aggregation}, where variations in coding---e.g., many small drifts in different kinds of heart failure that could add up to a large drift in heart failure at a higher level of representation---could normally cause drift to be overlooked (R5). In addition to areas where drift increases (Figure \ref{fig:hierarchy_difference}a), areas where drift decreases (Figure \ref{fig:hierarchy_difference}b) may also help define areas where drift has aggregated, by indicating smaller drifts that may have accumulated. 

In general, patterns of drift through a hierarchy can be varied, but areas where the amount of drift \textit{changes} substantially can be suggestive of informative portions of the hierarchy with respect to detecting the presence of
selection bias. In order to identify such areas, when comparing two cohorts we explicitly compute the difference in drift value---the drift gradient---between each child dimension and parent dimension in the hierarchy. Based on the selection bias distance metric (Equation \ref{eq:hellinger}), we define the drift gradient between a child dimension $d_c$ and its parent $d_p$ for two cohorts $c_i$ and $c_j$ as:
\begin{equation}
    \Delta H(d_c, d_p) = H(d_{ic},d_{jc}) - H(d_{ip},d_{jp})
\end{equation}

To capture (1) areas of the hierarchy with large increases in drift and (2) areas of the hierarchy with large decreases in drift, we define a gradient-based saliency criterion for a dimension $d_i$, given a single parent dimension $d_p$, and a set of child dimensions $D_c = \{ d_{cj} \}$ (empty for leaf nodes):
\begin{equation}
    S(d_i) = \Delta H(d_p,d_i) \geq t_s \ \lor \ \exists d_{cj} \in D_c \ -\Delta H(d_i,d_{cj}) \leq {-t_s} \ ,
\label{eq:saliency}
\end{equation}
where $t_s$ is a user-specified threshold. This equation detects whether a given node has increased (Figure \ref{fig:hierarchy_difference}a, \textcolor{OliveGreen}{green}) or any of its children have decreased (Figure \ref{fig:hierarchy_difference}b, \textcolor{BrickRed}{red}), by an amount $\geq t_s$. 
The user can adjust $t_s$ to determine the degree of aggregation used for the hierarchical visualization methods described in Sections \ref{sec:IciclePlot} and \ref{sec:DotPlot}. 




\subsection{Constraints}
\label{sec:Constraints}

Typically, the dimensions included in the filter constraints used to define cohorts will exhibit the greatest amount of drift. For example, if a user derives one cohort from another by constraining by \textit{Gender = Female}, the \textit{Gender} dimension would, by design, be expected to exhibit a very large drift in distribution. However, when tracking selection bias, the goal is to convey the magnitude of \textit{unintended side effects}. 

Previous work has dealt with this issue by simply excluding all constrained dimensions from consideration when tracking or visualizing selection bias \cite{gotz_adaptive_2016,gotz_adaptive_2017}. 
However, when filtering using hierarchically related dimensions, this naive approach is not sufficient.  In particular, large drifts in distribution can often occur for the descendants of a constrained dimension. Moreover, it can be assumed that the drift in descendant dimensions is expected by the user. If not, the user could have filtered using dimensions at a lower and more specific level in the hierarchy. We therefore exclude from the calculation of $H_{avg}$ both: (1) dimensions explicitly referenced by a constraint, and (2) any descendants of the explicitly referenced dimensions.

In contrast, when \emph{visualizing} these metrics in detail, excluding constrained dimensions is not desirable because the drift in distributions in their descendant dimensions may be informative to users. Rather than removing constrained dimensions from our hierarchical visualizations, we instead implement two design choices: (1) indicating visually which dimensions are constrained, and (2) normalizing scales to reduce the extent to which constrained variables and their descendants dominate the visualization. For (1), in all detailed cohort comparison visualizations (Sections \ref{sec:IciclePlot}-\ref{sec:ListView}), constraints are indicated with a diamond symbol $\blacklozenge$. For (2), all color scales are normalized to the maximum distance for non-constrained and non-constrained-descendant dimensions. In this manner, the constraints remain part of the visualization---enabling the user to see variations in drift between constrained descendants---while still enabling the user to effectively find non-constrained dimensions that may be subject to selection bias.

\subsection{Baseline and Focus}

Selection bias occurs when a cohort is selected in such a way that proper randomization (of the non-constraining dimensions in the data) is not conducted, making the cohort no-longer representative of the larger population intended to be analyzed. This larger population can be thought of as a \textit{baseline} against which cohorts can be compared to check for potential selection bias. During a given analysis, it may be useful to explore different baselines for comparison. It is therefore necessary to support a \textit{flexible} approach for choosing the current baseline. In the cohort provenance tree (Section \ref{sec:CohortTree}), the initial queried dataset serves as the baseline by default, however the user can select any created cohort to serve as the baseline against which all other cohorts are compared. 

In addition, to enable more in-depth comparison of any given cohort against the baseline, we adopt the concept of a \textit{focus} cohort. By default the focus cohort is the most recently created, however the user can select any cohort to serve as the focus. The visualizations supporting in-depth comparison of the potential selection bias between the focus and baseline cohorts are described in Sections \ref{sec:IciclePlot}-\ref{sec:VariableDistribution}.

%% file: sections/design.tex
\section{Visualization Design and Implementation}

The visual interfaces for the selection bias tracking and cohort comparison features in \emph{Cadence} are motivated by the design requirements from Section~\ref{sec:Requirements} and build upon the selection bias measures presented in Section~\ref{sec:measuringbias}. 
The Cohort Tree panel (Figure \ref{fig:cadence}a) contains the cohort provenance and selection bias tracking visualizations, which address R1, R2, and R3. The Cohort Comparison panel (Figure \ref{fig:cadence}, b-d) enables in-depth comparison of the current baseline and focus cohorts, addressing R4 and R5. It consists of a cohort overlap visualization (Section \ref{sec:CohortOverlap}), three tabbed views for in-depth exploration of the degree of drift across all dimensions in the dataset (Sections \ref{sec:IciclePlot}-\ref{sec:ListView}), and a detailed univariate comparison view that provides data-type-dependent visualizations to compare the distributions for a single dimension (Section \ref{sec:VariableDistribution}).

\subsection{Cohort Provenance Tree}
\label{sec:CohortTree}

\begin{figure}
    \centering
    \includegraphics[width=0.5\textwidth]{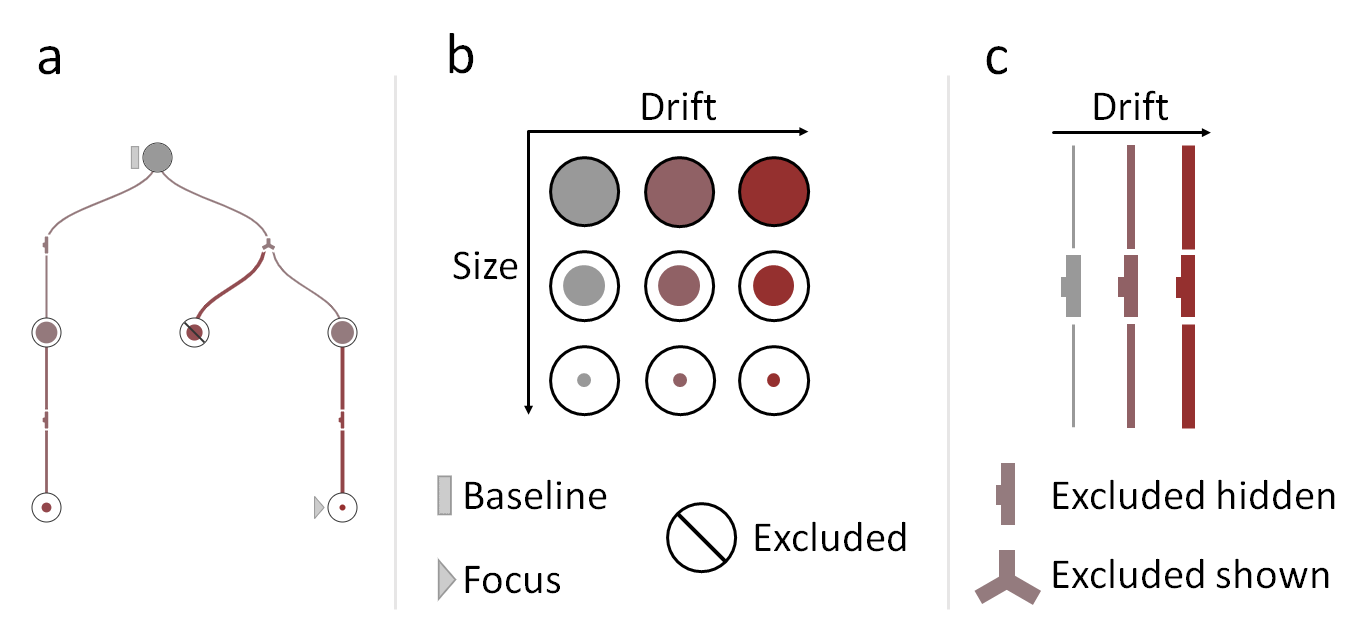}
    \caption{Cohort provenance visualization and iconography. (a) Each cohort and filter operation is shown as a node-link tree diagram. (b) Cohort glyph inner circle area is proportional to cohort size. Drift is encoded by a grey-red color map. The current baseline and focus cohorts are indicated by rectangular and triangular icons. Excluded cohorts that have been made visible are indicated by a diagonal slash. (c) Link color and thickness, and filter glyph color, indicate the amount of drift introduced at that filter step. Glyph appearance changes to indicate excluded cohort visibility.}
    \label{fig:cohort_tree}
\end{figure}

\begin{figure*}[t]
    \centering
    \includegraphics[width=0.95\linewidth]{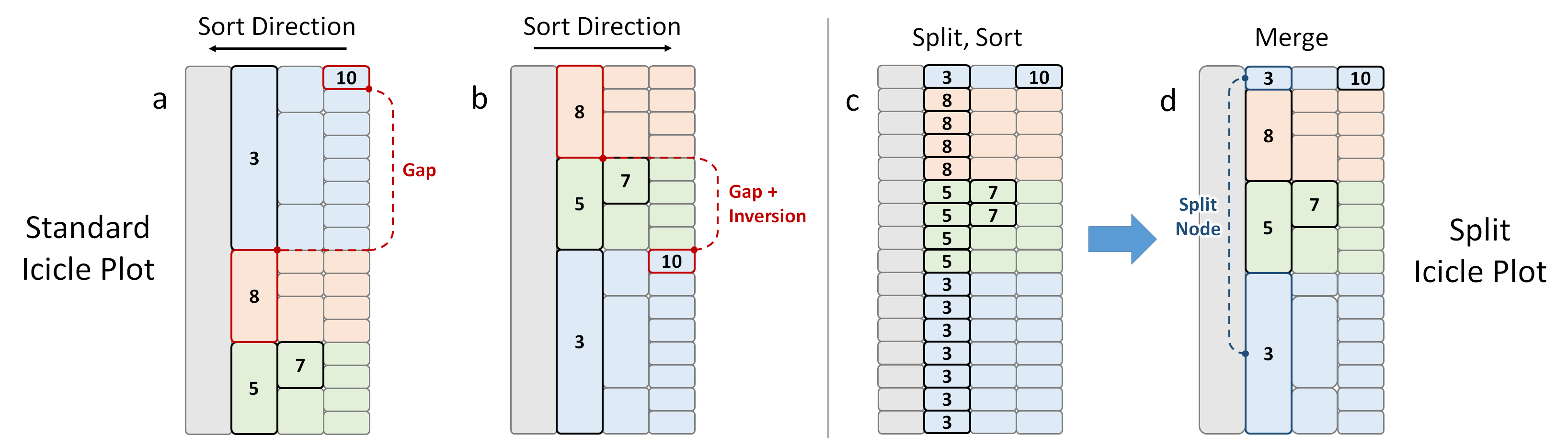}
    \caption{Sorting  by descending order in an icicle plot. The highest 5 values are displayed. With a standard icicle plot (left), recursive sorting strategies can be used to sort (a) from leaf to parent or (b) from parent to leaf. However, nodes may not be correctly ordered, and nodes with high values can become ``buried'' far from the top, resulting in \textcolor{BrickRed}{gaps} and \textcolor{BrickRed}{inversions}. With the split icicle plot layout (right), (c) the path from each leaf node to the root is first separated, and then sorted based on the maximum value along that path. (d) Adjacent nodes are then merged back together. For a given node, all paths above it will contain a value greater than or equal to its value, avoiding issues of gaps and inversions. However, some nodes may remain \textcolor{NavyBlue}{split} to achieve this guarantee.}
    \label{fig:icicle_combo}
\end{figure*}

To address R1 (Section \ref{sec:Requirements}), we show the provenance of each cohort created by the user via the \textit{Cadence} interface in a node-link tree diagram (Figure \ref{fig:cohort_tree}). Thisdesign is influenced by CONSORT diagrams (Figure \ref{fig:flow_diagram}), which are familiar to many medical experts in the target user population for \textit{Cadence}. A tidy-tree \cite{reingold_tidier_1981} provides a compact layout while showing each step in the cohort-creation process at a distinct level. Each node in the tree represents a cohort, with links representing dependencies (e.g., new cohort B was derived from previous cohort A).  

Nodes are represented by circular glyphs of unit size.  The glyphs include an inner circle with area proportional to the number of patients in the cohort (Figure~\ref{fig:cohort_tree}b).  The inner circles are color-coded based on $H_{avg}$ (Equation \ref{eq:hellinger_avg}), as computed by comparing the glyph's cohort to the current baseline. This design supports R2.

The links between cohorts include glyphs representing the filter operations used to derive each link's corresponding new cohort. The filter glyphs and the links themselves are both color-mapped by $\Delta H_{avg}$, the difference in $H_{avg}$ between the two cohorts that the link connects.  In this way, the edges visualize the amount of drift introduced directly by the link's corresponding filter operation.  The encoding is shown in Figure~\ref{fig:cohort_tree}c. Mousing over the filter glyph displays a tooltip with information about the constraints applied at that step in the cohort creation process, as well as the amount of drift introduced in response.  This design supports R3.

Critically, each filter operation creates both a set of included patients (those matching the filter criteria), and a set of excluded patients (those not matching the criteria). By default only the included cohort is shown. However, the user can show any excluded cohort via a context menu available for each filter glyph. This feature can help users understand the effect of a filter (as evidenced by the specification of excluded groups within the CONSORT flow diagram).  Excluded cohorts are depicted with a diagonal slash across the cohort glyph.  Moreover, a filter glyph's appearance changes to indicate that the excluded cohort is being shown (Figure~\ref{fig:cohort_tree}c). 

The cohort tree also indicates the currently selected baseline and focus cohorts. The baseline is indicated with a rectangular icon to the left of the baseline cohort glyph, and the focus with a triangular icon (Figure~\ref{fig:cohort_tree}b). By default, the first cohort used at the start of a visual analysis session is marked as the baseline.  Meanwhile, the focus is updated (by default) to reference each new cohort as it is created. However, users are able to interactively select any cohort in the tree at any time to serve as the baseline or focus through a popup context menu available for each cohort glyph. This functionality provides control over which cohorts are the baseline and focus used for the detailed cohort comparison visualizations described in the remainder of this section.

\subsection{Cohort Overlap}
\label{sec:CohortOverlap}

\begin{figure}
    \centering
    \includegraphics[width=0.4\textwidth]{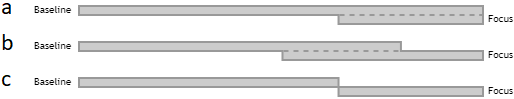}
    \caption{The cohort overlap visualization compares the focus and baseline cohorts, showing the relative sizes of the cohorts and proportion of individuals that belong to both. This includes cases where (a) the focus is a subset of the baseline, (b) the two cohorts partially overlap, and (c) the cohorts are disjoint.}
    \label{fig:cohort_overlap}
\end{figure}

The first element of the Cohort Comparison panel, which enables users to compare the baseline and focus cohorts, is the cohort overlap visualization.  This view depicts the proportion of patients that are members of the baseline and focus cohorts.  Unlike previous work with a linear provenance model, in which the focus was restricted by definition to be a subset of the baseline (e.g. \cite{gotz_adaptive_2016,gotz_adaptive_2017}), our tree-based approach enables a more flexible set of comparisons.  As a result, there are three possible overlap relationships to be visualized. Figure \ref{fig:cohort_overlap} shows examples of these three conditions: (a) one cohort (typically the focus) is a subset of the other (typically the baseline), (b) there is partial overlap, and (c) the two cohorts are disjoint. This visualization conveys the type of overlap the two cohorts have, the relative size of the cohorts, and the proportion of each cohort that falls within the overlapping subset.  

\subsection{Split Icicle Plot}
\label{sec:IciclePlot}

A key element of the visual design for comparing cohorts is to help users identify where in the dimension hierarchy the largest drifts in distribution have occurred. Given the large hierarchies in our data (Table \ref{tab:querystats}), our initial attempts to visualize this information employed tree maps \cite{shneiderman_tree_1992} and icicle plots \cite{kruskal_icicle_1983}, which are both space-filling techniques with compact representations. However, despite their compact representations, the number of dimensions in the hierarchies made them too big to fit on screen without over-plotting issues. 

Attempting to remedy this problem, we first implemented a scrollable icicle plot, which revealed another problem: the inability of icicle plots (and other space-filling hierarchical representations) to enforce a strict ordering of dimensions by value. Icicle plots partition space into a rectangular node for each dimension, with each node placed adjacent to its parent along one axis and sized along the other axis with a length proportional to the number of descendant leaf nodes. Recursive strategies for ordering sibling nodes are possible (e.g., Figure~\ref{fig:icicle_combo}, a and b). However, nodes can become ``buried'' by the hierarchical structure of the visualization, leading to \textit{gaps} and \textit{inversions}. Figure~\ref{fig:icicle_combo}a shows how a leaf-to-root sorting strategy can produce a large gap between the largest (10) and 2\textsuperscript{nd} largest (8) nodes by value. Similarly, Figure~\ref{fig:icicle_combo}b shows how a root-to-leaf sorting strategy can produce both an inversion and a gap between nodes that are neighbors in sort order.

This issue is problematic for visualizing drift in large hierarchies for two reasons: (1) ordering is an important cue for understanding which dimensions have shifted the most, and it is useful for users to be able to scan from top-to bottom to identify areas of potential selection bias, and (2) when using a scrollable interface to avoid over-plotting, users will need to scroll to find dimensions with large amounts of drift if they are not ordered with largest drift first.

\subsubsection{Algorithm and Example}

To address these issues, we developed the split icicle plot (Figure~\ref{fig:icicle_combo}, c and d). The split icicle plot algorithm proceeds in three stages: \textit{split}, \textit{sort}, and \textit{merge}. The hierarchical data are first \textit{split} into paths from each leaf to the root. The paths are then \textit{sorted} by the maximum value along each path. Adjacent nodes are then \textit{merged} back together. The result is a layout in which, for a given node in the visualization, all paths to the root above it will contain a value greater than or equal to its value, avoiding issues of gaps or inversions. As a tradeoff however, some nodes may be ``split'' to achieve this guarantee (Figure~\ref{fig:icicle_combo}d). 

\begin{figure}[t]
    \centering
    \includegraphics[width=0.5\textwidth]{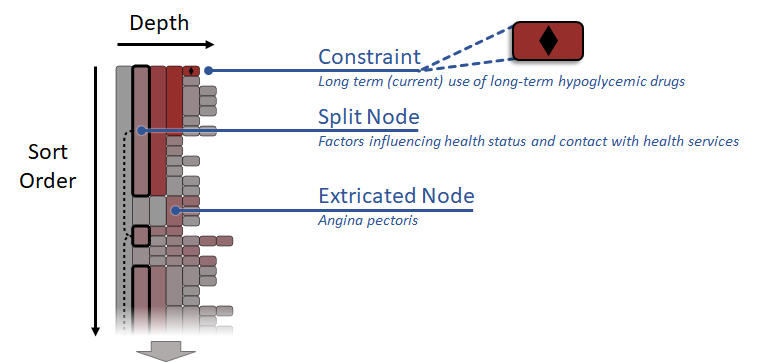}
    \vspace{-0.5cm}
    \caption{Split icicle plot visualizing drift between two cohorts in the ICD-10-CM hierarchy. Drift is encoded with a grey-red color map, and constraints are indicated with a $\blacklozenge$ symbol. Split nodes are indicated by dashed lines joining the split sections of the node on mouseover. A split ancestor of the constraint enables an otherwise buried node with high drift to be \textit{extricated}.}
    \label{fig:split_icicle_plot}
    \vspace{-0.2cm}
\end{figure}

Figure~\ref{fig:split_icicle_plot} shows an actual example of the split icicle plot visualizing differences in ICD-10-CM diagnoses between two cohorts. In addition to sorting the leaf-to-root paths in descending order by maximum drift (placing largest drift values at the top), the drift for each individual variable in the hierarchy is encoded using a grey-red color map, scaled to the non-constrained variable with the most drift. The one constrained variable in this example is indicated with a $\blacklozenge$ symbol, as explained in Section \ref{sec:Constraints}. Split nodes are indicated by dashed lines that join the split sections on mouseover. In this example, a parent node of the constraint has been split, enabling a variable that would otherwise have been buried below this node to be \textit{extricated} and revealed as the most distant variable that is not an ancestor of the constraint. Split nodes ``break'' the strict hierarchical layout traditionally imposed by icicle plots, which could potentially hinder interpretation. However in practice the node splits tend to form subgroups of children with similar drift values which aids the user's comprehension of the visualization.


\subsubsection{Hierarchical Aggregation}

\begin{figure}[t]
    \centering
    \includegraphics[width=0.5\textwidth]{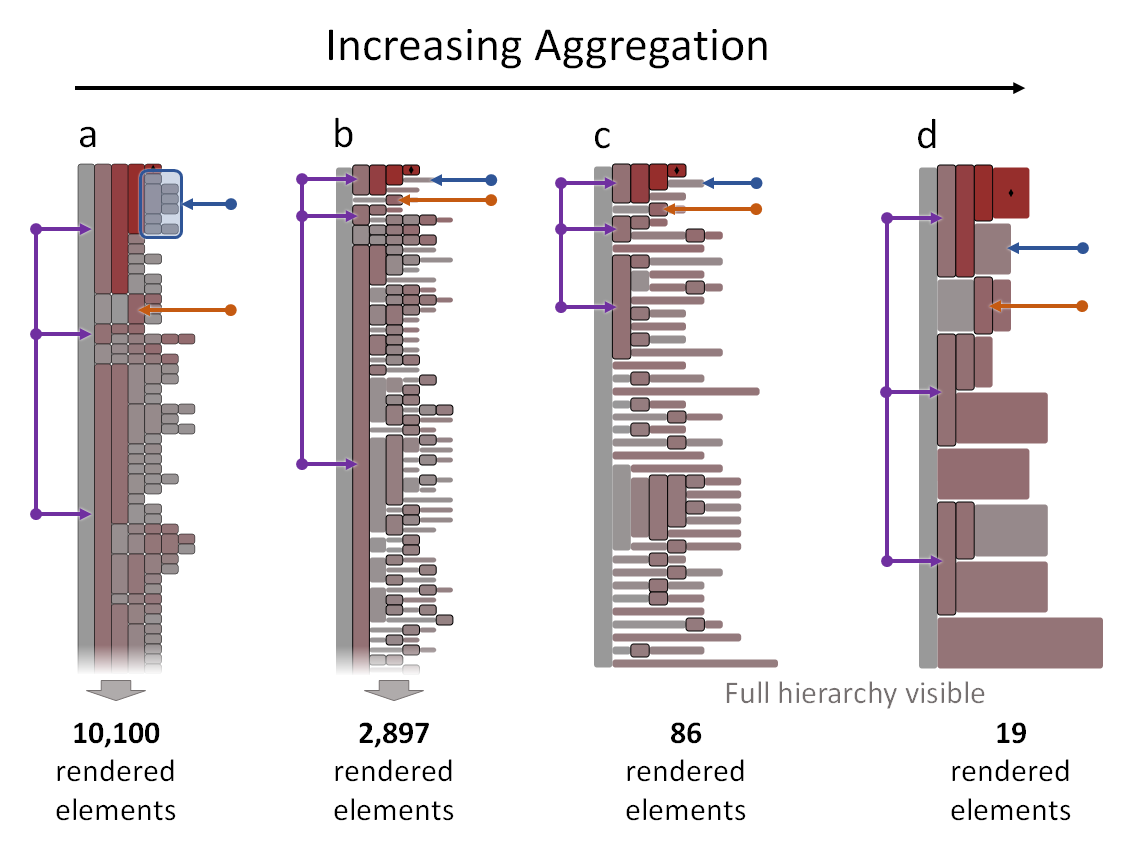}
    \vspace{-0.5cm}
    \caption{Hierarchical aggregation using gradient-based saliency. Nodes that do not meet the saliency criterion (Equation \ref{eq:saliency}) are merged into groups. Nodes that do meet the criterion are outlined in black. The \textcolor{NavyBlue}{blue} arrows indicate a group of nodes that is merged from (a) to (b), maintained from (b) to (c), and then merged with another group from (c) to (d). The \textcolor{Bittersweet}{orange} arrows indicate the same \textit{extricated} node from Figure \ref{fig:split_icicle_plot}, which is maintained from (a-d), as is the split node from the same figure, indicated in \textcolor{Violet}{purple}.} 
    \label{fig:aggregation_threshold}
    \vspace{-0.15cm}
\end{figure}

The split icicle plot solves issues related to over-plotting and ordering, and provides an effective visualization of drift in hierarchical data. However, two issues remain when this approach is applied to the large hierarchies found in our application: (1) drawing individual nodes for each dimension in a large hierarchy can hinder performance during interaction, and (2) the large size of the visualization and the use of scrolling (to help overcome over-plotting) can make it difficult for users to obtain a broad overview of the data. To address these issues we have developed a hierarchical aggregation technique with a user-controlled level of aggregation to simplify the visual representation (Figure~\ref{fig:aggregation_threshold}). 

\begin{figure}[t]
    \centering
    \includegraphics[width=0.34\textwidth]{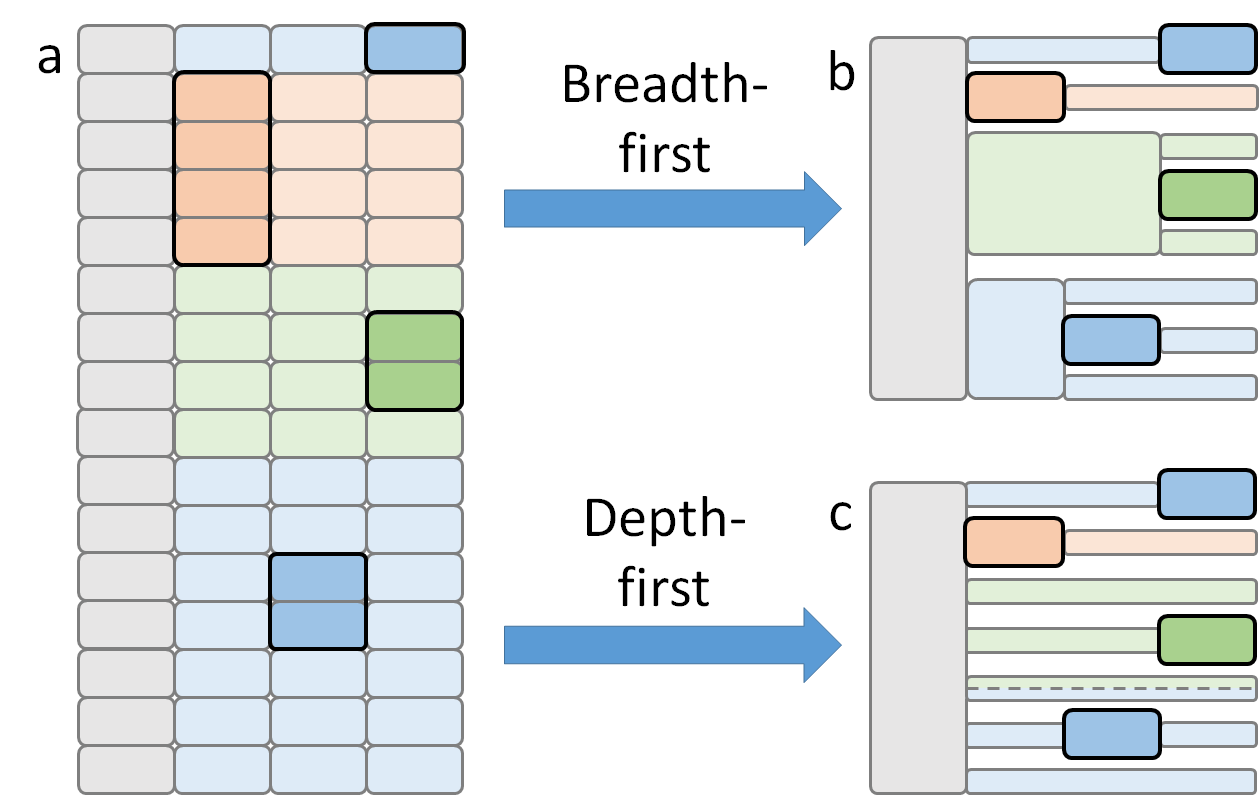}
    \caption{Two hierarchical aggregation methods for split icicle plots. Given (a)~a split, sorted, and unmerged icicle plot with salient nodes outlined in black, (b)~breadth-first aggregation prioritizes the merging of adjacent non-salient nodes within the same level of the hierarchy. This preserves much of the hierarchical structure. (c)~Depth-first aggregation, which can be more efficient for deep hierarchies, prioritizes the merging of non-salient nodes along each path from the root to a leaf.} 
    \label{fig:aggregation_illustration}
\end{figure}

The aggregation algorithm relies on gradient-based saliency as defined in Equation~\ref{eq:saliency}. 
Beginning with a split, sorted, and unmerged icicle plot (Figure \ref{fig:icicle_combo}c), we have developed two aggregation methods: \textit{breadth-first}, and \textit{depth-first}.   Figure~\ref{fig:aggregation_illustration} demonstrates both methods. Breadth-first aggregation merges adjacent nodes at the same level in the hierarchy if they either (1) share a common salient descendant, or (2) have no salient descendants. These merged groups are propagated down the tree until a salient node or leaf node is encountered, at which point new groups are created. In constrast, depth-first aggregation merges nodes along each path from the root until either (1) a salient node is encountered, at which point new paths are created for any children, or (b) a leaf node is encountered. Any groups that share the same starting or ending point are then merged together. 
In practice, breadth-first tends to preserve more of the hierarchical structure, as adjacent nodes for the same dimension are merged. In contrast, depth-first is more likely to split nodes for the same dimension, but can be more efficient in aggregating hierarchies with large depths. The examples used in this paper all use breadth-first aggregation.

\begin{figure}[t]
    \centering
    \includegraphics[width=0.5\textwidth]{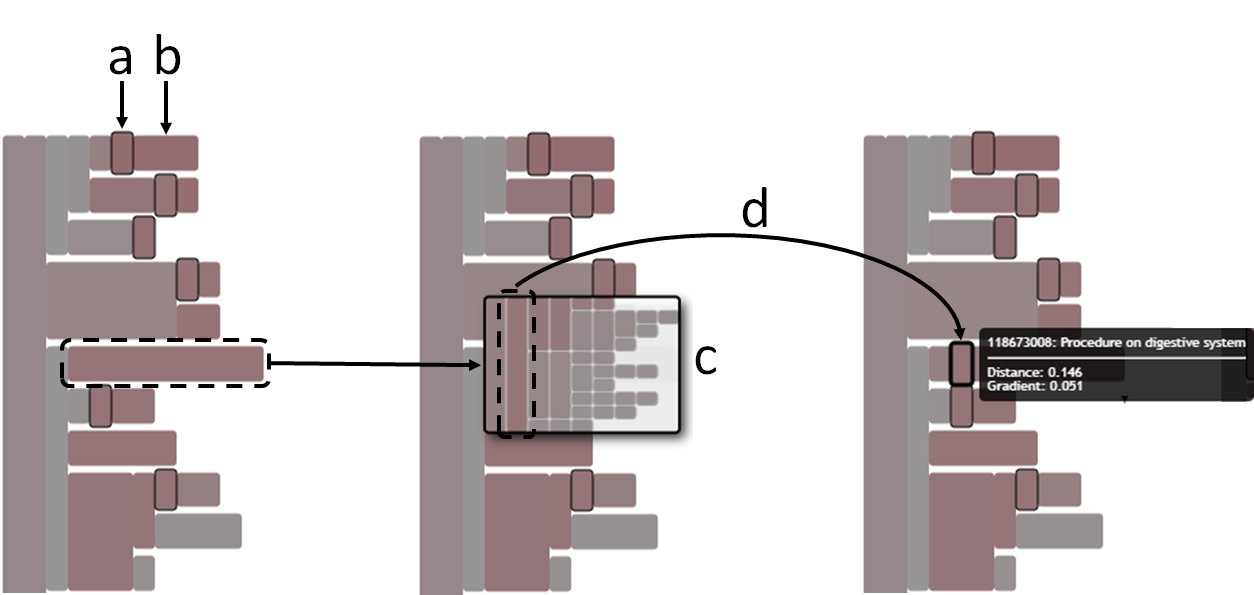}
    \caption{Split icicle plots with hierarchical aggregation. (a)~Salient nodes are drawn with black outlines. (b)~Merged groups are drawn with reduced heights. A grey-red color map indicates drift. For groups, the maximum drift in the group is used. 
    Placing the cursor over a group will cause it to expand~(c), showing a standard icicle plot of the nodes in that group. The user can select a nodes in the expanded group, and the layout will be updated with that node manually defined as a salient node~(d).} 
    \label{fig:aggregation_example}
    \vspace{-0.15cm}
\end{figure}

Figure~\ref{fig:aggregation_example} shows an example of the split icicle plot with hierarchical aggregation. Salient nodes are drawn with black outlines (Figure~\ref{fig:aggregation_example}a), whereas merged groups are drawn with reduced heights to further emphasize the salient nodes (Figure~\ref{fig:aggregation_example}b). Each node or group is colored by drift with a grey-red color map scaled to the maximum drift for a non-constrained dimension.
For groups, the color is determined by the maximum drift in the group
to indicate areas where large drift may have occurred that did not meet the current saliency definition, such as a chain of relatively small increases in drift that combine to produce a large total drift. The user can inspect the contents of each group with the cursor, which will display a standard icicle plot of the nodes contained within the group (Figure~\ref{fig:aggregation_example}c).
The user can select a node from within the expanded group to manually define the corresponding dimension as salient.  In response, the overall layout will be updated accordingly.


\subsection{Hierarchical Dot Plot}
\label{sec:DotPlot}

One drawback of the split icicle plot visualization is that it depends on color to encode drift, which can make it difficult to distinguish smaller value differences. We therefore developed a hierarchical dot plot visualization as an alternative that uses position to encode drift (Figure~\ref{fig:teaser}d).


As with the split icicle plot, hierarchical aggregation is used to reduce over-plotting and increase rendering performance using the same gradient-based saliency method. The visual design is similar to that of Figure \ref{fig:hierarchy_difference}. Salient nodes are rendered as dots, with x-position representing the depth of the node in the hierarchy, and y-position the amount of drift. The size of the node is proportional to $\lvert \Delta H(p,e) \rvert$, and $\Delta H(p,e)$ is mapped to a blue-grey-red color map, with blue for negative values and red for positive values. The user can highlight any visible node via mouseover to show links to all of its ancestors and descendants. Nodes below the saliency threshold are aggregated and displayed as a background heat-map. Any heat-map cell can be expanded to reveal the nodes inside, and those interior nodes can themselves be selected. By iteratively selecting ancestor/descendant nodes, the user can interactively explore the hierarchy (addressing R4 and R5). This dot plot design provides better visual fidelity for drift values when compared to the split icicle plot, but the overall hierarchical structure is less apparent.

\subsection{List View}
\label{sec:ListView}

In addition to the hierarchical visualization techniques described above, we also provide a list view that shows a basic table of dimensions in descending order by drift.  The table includes a bar chart representation of the drift values as shown in Figure \ref{fig:teaser}c. This view ignores the hierarchical relationships in the data, but has been provided as a simple, easy-to-use format that is familiar to users.

\subsection{Variable Distribution}
\label{sec:VariableDistribution}

The in-depth cohort comparison visualizations listed above enable the user to see which dimensions in the hierarchy have drifted, and the magnitudes of those drifts. However, they do not convey $how$ the dimensions differ between the two cohorts. For example, filtering for patients with a SNOMED-CT code for \textit{Procedure on hip} reveals a large drift in distribution for the ICD-10-CM code for \textit{Pain in hip}. In this case, it can be assumed that the cohort with a hip procedure also has a higher incidence of hip pain, but such relationships may not always be so obvious. We therefore enable the user to select any attribute or event to see a detailed visualization of the distributions of the selected dimension for each cohort.

Each of the detailed cohort comparison visualizations (split icicle plot, hierarchical dot plot, list) enable users to select 
any dimension from within the visualization by direct selection, or via explicit search with a search box. The dimension selection is linked across all views, enabling users to pivot between the different visualizations.

Moreover, once a dimension has been selected, a data-type-dependent visualization of the distributions for that dimension in both the baseline and focus cohorts is displayed at the bottom of the cohort comparison panel. As shown in Figure~\ref{fig:teaser}e, this distribution comparison view supports bar charts for categorical data, a histogram for numeric data, and horizontal binary bar charts for binary variables. In all cases, the proportion of each value is shown to enable comparison between cohorts of different sizes. Interactive highlighting shows contextual information on individual data values, such as the specific number of patients. These visualizations further R4 by enabling detailed distribution comparisons for individual variables.


%% file: sections/evaluation.tex
\section{Example Use Case and Domain Expert Interviews}

To demonstrate the benefits of the approaches outlined in this paper, we present an example use case for the selection bias tracking capabilities within \emph{Cadence}.  We also report a thematic analysis of qualitative feedback gathered through interviews with medical domain experts.

\subsection{Example Use Case}
\label{sec:UseCase}

\begin{figure*}
    \centering
    \includegraphics[width=0.88\linewidth]{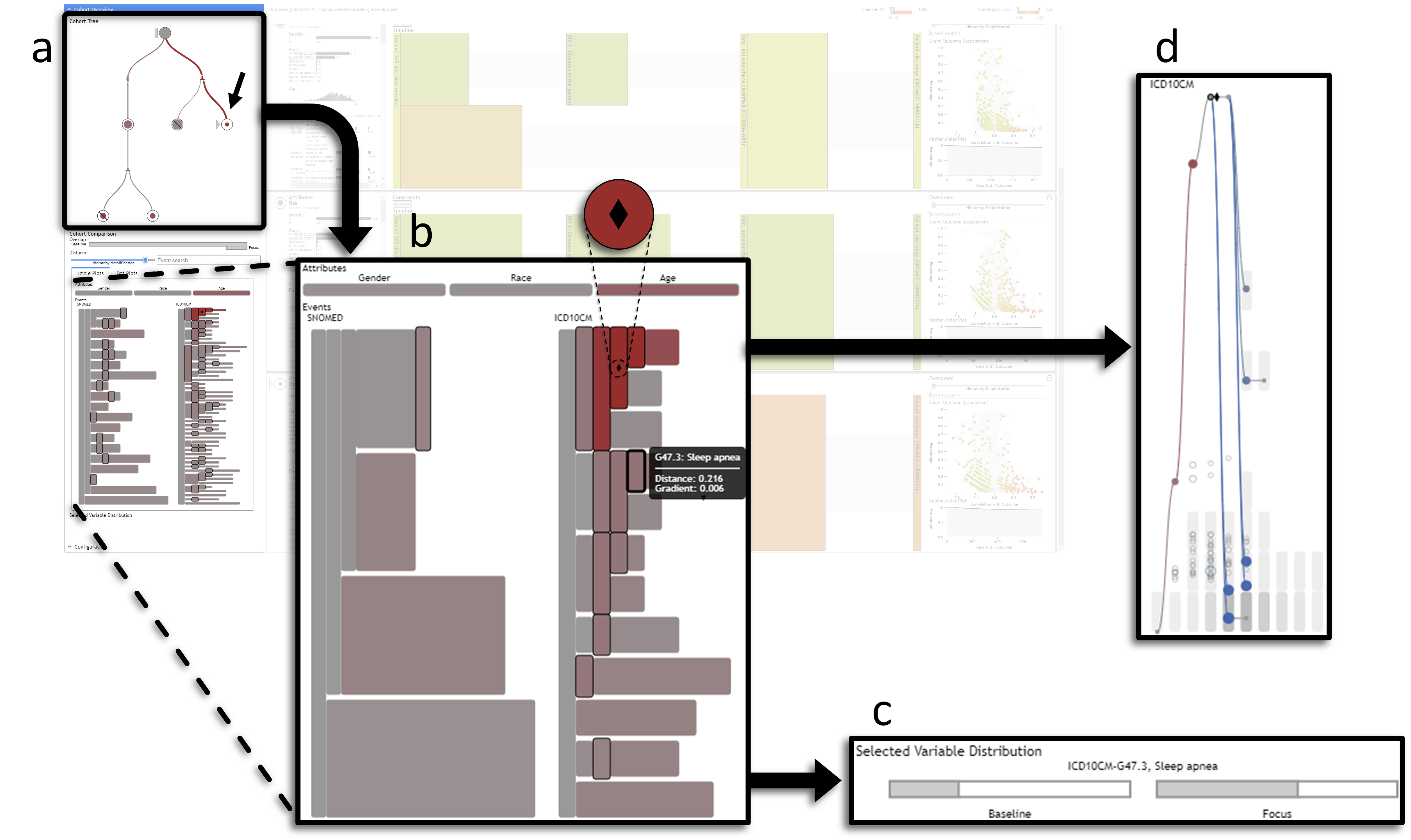}
    \caption{The example use case described in Section \ref{sec:UseCase}. The temporal event sequence visualization, which is not the focus of this paper, is faded.} 
    \label{fig:use_case}
\end{figure*}

Figure \ref{fig:use_case} shows a series of screenshots from \textit{Cadence} during an analysis of a cohort of 1,732 patients who were discharged from a hospital after having been diagnosed with some form of pain. The queried data includes one year's worth of medical event data (SNOMED-CT and ICD-10-CM codes) prior to the pain diagnosis, as well as data from the period between the pain diagnosis and the hospitalization.

Using the temporal event analysis portion of \textit{Cadence} (faded in Figure~\ref{fig:use_case}), the user has created a number of cohorts based on filters that reference various attributes and events surfaced by the interface.  Examining the cohort tree, the user notices that the current focus cohort of 227 patients, filtered by \textit{Obesity and other hyperalimentation}, seems to have drifted considerably farther than other cohorts from the baseline (in this case the initial query result). The high drift is indicated by the red color of the path to the cohort and the cohort glyph itself in Figure~\ref{fig:use_case}a. The user decides to investigate the differences between the focus and baseline cohorts. After adjusting the aggregation level for the split icicle plot (Figure~\ref{fig:use_case}b), the user identifies areas of the dimension hierarchy that are contributing the most to the high drift. As expected, the constrained \textit{Obesity...} variable, indicated by a $\blacklozenge$ symbol, has drifted the most. Moving on, the user highlights the ICD-10-CM event with the highest drift in a different branch of the hierarchy and finds it to be \textit{Sleep apnea}. Suspecting that patients in the \textit{Obesity...} cohort probably have a higher prevalence of \textit{Sleep apnea}, the user checks this assumption with the selected variable distribution visualization (Figure~\ref{fig:use_case}c). This reveals that the focus \textit{Obesity...} cohort contains a 59\% incidence rate (133 of 227 patients) of \textit{Sleep apnea}, whereas the baseline cohort contains just 29\% (496 of 1,732 patients).
Without selection bias tracking and detailed cohort comparison, the user would likely not notice that \textit{Sleep apnea} is more prevalent in the focus dataset than in the baseline. Having access to this information, however, enables the user to incorporate this information into any subsequent analyses of the data. For example, they could control for sleep disorders when studying the effectiveness of a particular medication for obese patients to ensure analytical validity for the target population.

Looking back at the split icicle plot, the user notices that some descendant nodes of the constrained variable also have large drift values (red), whereas others have a very low drift value (grey). The user selects the constrained variable and switches to the hierarchical dot plot view (Figure~\ref{fig:use_case}d) to inspect further. The user notices that four descendants maintains relatively high drift, whereas the drift for five other descendants drop precipitously. 
This pattern may indicate different subgroups of obesity warranting additional study.

\subsection{Domain Expert Interviews}
 
To better understand the strengths and weaknesses of the selection bias tracking capabilities outlined in this paper, we conducted a set of semi-structured interviews with three medical experts.

All three participants were health-focused researchers with data analysis experience.  One was a medical doctor with both clinical and research responsibilities, whereas both of the others were PhD-level researchers.  All three participants were full-time university employees with experience using the i2b2 system \cite{murphy_serving_2010}, an NIH-funded cohort selection platform that has been deployed to support data-driven health research at a large number of research universities.

Each participant met individually with two study moderators for a one hour session. Participants were first introduced to \emph{Cadence} and provided a demonstration of cohort selection tasks using the system.  After the demonstration, participants were asked questions in a semi-structured interview format.  The interviews triggered additional interactions with the visualization system.

\subsubsection{Thematic Analysis of Interview Findings}

The three domain experts provided feedback about many aspects of the \emph{Cadence} visual analytics system.  We summarize their comments with a thematic analysis of the interview findings, focusing on the feedback that is most relevant to the selection bias tracking and cohort comparison capabilities.

{\bf General feedback.} General comments about the approach were typically very positive, and participants agreed that the system provided many clear benefits. One participant commented that it enables you to ``instantly generate insights'' and is a powerful ``hypothesis generating application.'' Other feedback included, ``I really like your design,'' this is a ``really powerful analytical tool,'' and ``this is very useful...for cohort studies.'' Although feedback was positive, the participants did acknowledge the complexity of the system, and the need for training. One participant noted that it seemed complicated at first glance, but that ``it made sense'' after seeing it demonstrated. Another participant noted that some degree of complexity and training should be expected: ``i2b2 was complicated at first,'' but that such tools ``are for skilled users'' and require training. To sum up, one participant stated that this is ``very cool, but there is a learning curve.''

{\bf Benefits of cohort tree and selection bias tracking.} Participants agreed that selection bias, and other issues related to the validity of analytical findings, such as ``fishing expeditions,'' are important, and germane to their work, e.g. ``that's something I'm really concerned about.'' However they did raise the issue of how to indicate ``how much [selection bias] is too much?'' Participants also agreed that the cohort tree is a useful method for keeping track of the provenance of created cohorts. One participant stated that ``it helps you not get lost. It's breadcrumbs.'' Another stated that ``if I get lost here, I can go back [to the cohort tree] to orient myself.'' Some suggestions for improvements included providing the ability to toggle baseline and focus cohort tooltips--instead of showing them only on mouseover--to serve as a reminder when interacting with the other parts of the visualization. This feature has since been implemented.

{\bf Benefits of detailed cohort comparison.} Participants saw benefits and drawbacks of all three detailed cohort comparison methods. Regarding the hierarchical visualizations, one participant remarked, ``this is good, but takes extra effort to see what it means,'' due to the need to mouseover to obtain detailed information on the event type. Another participant stated that ``All those views are very useful.'' Regarding a particular selected variable distribution visualization, one participant highlighted a discovery during the interview session: when we ``saw the spike in the age distribution, [I asked] why? The system lets you look into it right away.'' One participant wished to ``make the comparison view larger,'' suggesting that it would be useful to be able to switch between cohort selection and cohort analysis modes. 

{\bf Alternative visualization designs.} More specific feedback on the alternative visualization designs for the detailed cohort comparison visualizations included one participant's view that the icicle plot ``was good for comparison'' and that they intuitively understood the use of color. However, for the dot plot, the same participant suggested that ``when it is crowded, hard to see.'' In general, participants found that the hierarchical visualizations were ``good for exploring,'' whereas the list was ``good for finding specifics.''

%% file: sections/conclusion.tex
\section{Conclusion}

This paper presents a new selection bias tracking and high-dimensional comparison system for exploratory cohort selection. It overcomes limitations in prior work by (1)~using a tree-based cohort provenance system and (2)~leveraging hierarchical relationships between data dimensions to both better communicate areas of drift in variable distributions between cohorts, and enable user-controlled aggregation for improved visualization.


The approach is demonstrated by integration with a medical temporal event sequence visual analytics tool, and implemented via: (1)~a cohort provenance tree visualization that enables the user to keep track of created cohorts and indicates when selection bias may have occurred, (2)~a novel split icicle plot that improves the ability to order nodes by value in a hierarchical visualization, highlights areas of potential selection bias, and provides user-controlled aggregation, and (3)~a set of complementary visualizations, including a hierarchical dot plot, list view, and detailed variable distribution visualizations, to examine the differences between two cohorts in detail.

Feedback from domain expert interviews indicated that selection bias tracking in visual analytics tools is an important capability, and that further work in this vein is warranted. Specific areas for future work include: (1)~providing a reference for potentially ``dangerous'' amounts of selection bias that is comparable across particular analyses, (2)~exploring techniques to correct for selection bias implemented within the system, and (3)~investigating the selection bias tracking and cohort comparison techniques for applications beyond the medical domain. In addition, more in-depth evaluations of the \textit{Cadence} system, and its individual components, are planned.